\documentclass[useAMS,usenatbib]{mn2e}              
\usepackage{epsfig}
\def\reff@jnl#1{{\rm#1\/}}

\def\aj{\reff@jnl{AJ}}                  
\def\araa{\reff@jnl{ARA\&A}}            
\def\apj{\reff@jnl{ApJ}}                
\def\apjl{\reff@jnl{ApJ}}               
\def\apjs{\reff@jnl{ApJS}}              
\def\apss{\reff@jnl{Ap\&SS}}            
\def\aap{\reff@jnl{A\&A}}               
\def\aapr{\reff@jnl{A\&A~Rev.}}         
\def\aaps{\reff@jnl{A\&AS}}             
\def\baas{\reff@jnl{BAAS}}              
\def\jrasc{\reff@jnl{JRASC}}            
\def\memras{\reff@jnl{MmRAS}}           
\def\mnras{\reff@jnl{MNRAS}}            
\def\physrep{\reff@jnl{Phys.Rep.}}
\def\pra{\reff@jnl{Phys.Rev.A}}         
\def\prb{\reff@jnl{Phys.Rev.B}}         
\def\prc{\reff@jnl{Phys.Rev.C}}         
\def\prd{\reff@jnl{Phys.Rev.D}}         
\def\prl{\reff@jnl{Phys.Rev.Lett}}      
\def\pasp{\reff@jnl{PASP}}              
\def\pasj{\reff@jnl{PASJ}}              
\def\skytel{\reff@jnl{S\&T}}            
\def\solphys{\reff@jnl{Solar~Phys.}}    
\def\sovast{\reff@jnl{Soviet~Ast.}}     
\def\ssr{\reff@jnl{Space~Sci.Rev.}}     
\def\nat{\reff@jnl{Nature}}             

\title[Distribution of stellar mass] 
{The distribution of stellar mass in   the  low-redshift   Universe}  
\author[Li   \&   White]  
{Cheng Li$^{1,2}$\thanks{E-mail: leech@mpa-garching.mpg.de},     
 Simon D.~M.  White$^{1}$\\  
$^{1}$Max-Planck-Institute  for  Astrophysics,
      Karl-Schwarzschild-Str. 1, D-85741 Garching, Germany \\  
$^{2}$MPA/SHAO  Joint  Center  for  Astrophysical  Cosmology  at
      Shanghai Astronomical Observatory,  Nandan Road 80, 
      Shanghai 200030, China}

\begin{document}

\date{Accepted ........ Received ........; in original form ........}

\pagerange{\pageref{firstpage}--\pageref{lastpage}} \pubyear{2008}

\maketitle

\label{firstpage}

\begin{abstract}
We use a complete and uniform sample of almost half a million galaxies
from the Sloan Digital Sky  Survey to characterise the distribution of
stellar mass in the  low-redshift Universe. Galaxy abundances are well
determined over almost  four orders of magnitude in  stellar mass, and
are  reasonably but  not perfectly  fit by  a Schechter  function with
characteristic  stellar mass $m_\ast  = 6.7\times  10^{10}M_\odot$ and
with faint-end slope $\alpha = -1.155$. For a standard cosmology and a
standard stellar Initial  Mass Function, only 3.5\% of  the baryons in
the  low-redshift  Universe are  locked  up  in  stars. The  projected
autocorrelation  function of  stellar mass  is robustly  and precisely
determined for $r_p < 30h^{-1}{\rm Mpc}$. Over the range $10h^{-1}{\rm
  kpc}< r_p < 10h^{-1}{\rm Mpc}$ it is extremely well represented by a
power   law.   The  corresponding   three-dimensional  autocorrelation
function is $\xi^\ast(r)  = (r/6.1h^{-1}{\rm Mpc})^{-1.84}$.  Relative
to  the dark  matter, the  bias of  the stellar  mass  distribution is
approximately constant on large scales, but varies by a factor of five
for $r_p < 1h^{-1}{\rm Mpc}$.  This behaviour is approximately but not
perfectly  reproduced by current  models for  galaxy formation  in the
concordance $\Lambda$CDM cosmology.  Detailed comparison suggests that
a  fluctuation  amplitude  $\sigma_8\sim  0.8$  is  preferred  to  the
somewhat larger value adopted in the Millennium Simulation models with
which  we  compare  our  data.  This  comparison  also  suggests  that
observations  of  stellar  mass  autocorrelations  as  a  function  of
redshift might provide a powerful test for the nature of Dark Energy.
\end{abstract}


\section{Introduction}\label{S:introduction}

Four hundred years  ago Galileo turned his telescope  to the Milky Way
and discovered it to consist of countless faint stars. One hundred and
fifty  years later,  Kant speculated  that  it might  be an  enormous,
rotating stellar swarm,  held together by gravity in  a similar way to
the  Solar  System,  and  that  other nebulae  might  be  similar  but
extremely  distant  ``island universes''.   These  ideas were  finally
confirmed when Hubble established  the extragalactic distance scale in
the 1920's. Stars were accepted as  the dominant form of matter in the
Universe from this  time until the 1980's, when  new theoretical ideas
suggested that  the dark matter discovered  by \citet{Zwicky-33} might
consist     of    neutral,    non-baryonic     elementary    particles
\citep{Cowsik-McClelland-73, Peebles-82} and  X-ray images showed that
most  of  the  baryons in  rich  clusters  are  in  the form  of  hot,
intergalactic  gas \citep{Forman-Jones-82}.  It  now seems  clear that
baryons are not the dominant form  of matter in our Universe, and that
stars   account   for  only   a   small   fraction   of  the   baryons
\citep[e.g.][]{Fukugita-Hogan-Peebles-98,     Cole-01,    Komatsu-09}.
Nevertheless, stars are the only component of the cosmic mix for which
a complete and robust census  is possible. Such surveys teach us where
and with what efficiency baryons were converted into galaxies, and can
provide  stringent  constraints  on  our general  structure  formation
paradigm.

In  recent  years it  has  become clear  that  stellar  masses can  be
measured  for galaxies  in a  robust way  from  multi-band photometric
measurements     of     their     spectral    energy     distributions
\citep{Bell-deJong-01, Blanton-Roweis-07}  or from combined photometry
and spectroscopy \citep{Kauffmann-03}.  Low-mass stars contribute very
little to the light of galaxies, so the principal uncertainty in these
measurements comes  from the stellar Initial Mass  Function (IMF). For
any  particular  assumed  IMF,  the  uncertainties  due  to  dust,  to
metallicity, and to details of  the star-formation history turn out to
be  quite  small provided  reliable  photometry  is  available out  to
wavelengths of order 1 micron. Stellar mass is then a more natural way
to characterise the number of stars  in a galaxy than, for example, B-
or K-band luminosity.

In the current paper we use a complete sample of almost half a million
galaxies with excellent photometry and accurate redshifts to study the
distribution of stellar  mass  in  the low  redshift   universe.  This
separates into  two parts: the abundance of  galaxies as a function of
their stellar  mass,  and the clustering  of   stellar mass on  scales
larger than  those of individual  galaxies.  There  have been previous
studies of the first of these  statistics, the so-called mass function
of galaxies    \citep[e.g.][]{Cole-01,  Bell-03,  Wang-06,  Panter-07,
Baldry-Glazebrook-Driver-08}.   Our  results  agree with this previous
work, though with smaller statistical error bars because of the larger
sample (systematic uncertainties  due  to the IMF  remain as  large as
before, of course).

The  second  statistic  has  not,  to our  knowledge,  been  estimated
previously,  although there  have been  a few  measurements  of galaxy
clustering weighted by stellar {\it  light} or by {\it dynamical} mass
\citep[e.g.][]{Boerner-Mo-Zhou-89}.  As we show the autocorrelation of
stellar mass   is remarkable for the accuracy   with which  it  can be
estimated from our  sample, and for the fact  that it turns out  to be
almost a perfect  power law over three orders  of magnitude in spatial
scale.  The  near power-law   behaviour  of galaxy   correlations  was
emphasised in early work \citep[e.g.][]{Davis-Peebles-83} and has been
examined in some detail  in previous studies   with Sloan Digital  Sky
Survey data  \citep[e.g.][]{Zehavi-04,  Zehavi-05, Masjedi-06}.    The
latter  noted that deviations from  a pure power   law are detected at
high significance  in almost all  cases. In contrast, our stellar mass
autocorrelation does not  deviate from the  best-fit power law by more
than  12.5\% for  separations between  10~kpc  and 10~Mpc, a behaviour
which we will  show to be partially  but  not perfectly  reproduced by
existing galaxy formation models.

The structure of our paper is  as follows. In Section 2 we discuss the
observational dataset we analyse and the theoretical models with which
we compare  it.  Sections  3 and  4 then present  our results  for the
stellar mass function of galaxies and for the autocorrelation function
of stellar  mass, respectively.  A concluding  section discusses these
results, compares  them with model predictions, and  suggests that the
shape of  the mass autocorrelation  function might provide a  means to
estimate  how the  cosmic scale  factor and  the linear  growth factor
depend  on redshift,  and hence  to constrain  the properties  of Dark
Energy.

\section{Data}

\subsection{SDSS and NYU-VAGC}\label{sec:sample}

This    study    is    based     on    the    final    data    release
\citep[DR7;][]{Abazajian-08}   of  the   Sloan   Digital  Sky   Survey
\citep[SDSS;][]{York-00}. This contains images of a quarter of the sky
obtained using a  drift-scan camera \citep{Gunn-98} in the  {\em u, g,
  r, i, z} bands \citep{Fukugita-96,Smith-02,Ivezic-04}, together with
spectra of almost  a million objects obtained with  a fibre-fed double
spectrograph  \citep{Gunn-06}.   Both instruments  were  mounted on  a
special-purpose  2.5~meter telescope  \citep{Gunn-06} at  Apache Point
Observatory.      The     imaging     data     are     photometrically
\citep{Hogg-01,Tucker-06}    and    astrometrically    \citep{Pier-03}
calibrated, and were used to select spectroscopic targets for the main
galaxy  sample  \citep{Strauss-02},  the  luminous red  galaxy  sample
\citep{Eisenstein-01},  and  the  quasar  sample  \citep{Richards-02}.
Spectroscopic fibres  are assigned to  the targets using  an efficient
tiling     algorithm     designed     to     optimise     completeness
\citep{Blanton-03c}.  The details of  the survey strategy can be found
in \citet{York-00} and an overview  of the data pipelines and products
is provided in the Early Data Release paper \citep{Stoughton-02}. More
details on the photometric  pipeline can be found in \citet{Lupton-01}
and on the spectroscopic pipeline in \citet{Subbarao-02}.

For this  paper we take  data from {\tt  Sample dr72} of the  New York
University              Value              Added             Catalogue
(NYU-VAGC)\footnote{http://sdss.physics.nyu.edu/vagc/}.   This  is  an
update  of the  catalogue  constructed by  \citet{Blanton-05b} and  is
based on the full SDSS/DR7  data.  Starting from {\tt Sample dr72}, we
construct a magnitude-limited sample  of galaxies with $r\le 17.6$ and
spectroscopically measured redshifts in the range $0.001<z<0.5$.  Here
$r$  is  the  $r$-band  Petrosian apparent  magnitude,  corrected  for
Galactic  extinction, and the  apparent magnitude  limit is  chosen in
order to  get a sample  that is uniform  and complete over  the entire
area of the survey.  We also restrict ourselves to galaxies located in
the main contiguous  area of the survey in  the northern Galactic cap,
excluding the three  survey strips in the southern  cap (about 10\% of
the full survey area). These restrictions results in a final sample of
486,840 galaxies.

In  addition  to  the  magnitudes,  redshifts  and  positions  of  the
galaxies,  the NYU-VAGC  provides several  other quantities  which are
needed in our  analysis. The first is a stellar  mass for each galaxy,
which  is based  on its  redshift and  the five-band  SDSS photometric
data,  as  described  in  detail  in  \cite{Blanton-Roweis-07}.   This
estimate corrects implicitly for dust and assumes an universal Initial
Mass  Function  of \citet{Chabrier-03}  form.   As  we demonstrate  in
Appendix A, once all estimates are adapted to assume the same IMF, the
Blanton-Roweis masses  agree quite well  with those obtained  from the
simple, single-colour estimator of \citet{Bell-03} and also with those
derived by \citet{Kauffmann-03} from  a combination of SDSS photometry
and spectroscopy.  Given  the very large sample provided  by the SDSS,
sampling   fluctuations    and   ``cosmic   variance''    are   small.
Uncertainties in the mass estimation procedure dominate the systematic
error budget  for most  of the results  we present below.   Appendix A
shows  that such  uncertainties primarily  affect the  overall stellar
mass scale,  as do uncertainties in the  IMF itself, as long  as it is
assumed universal.  Results that depend  only on the  relative stellar
masses  of  galaxies  (for   example,  the  stellar  mass  correlation
function)  are therefore  much more  weakly affected  than  those that
depend directly on the mass  scale (for example, the mean stellar mass
density of the Universe).

The NYU-VAGC  also provides the  necessary information to  correct for
incompleteness in  our spectroscopic sample.  In particular,  we use a
mask which shows which areas of  the sky have been targeted, and which
have not, either because they are outside the survey boundary, because
they  contain   a  bright  confusing  source,   or  because  observing
conditions were  too poor to obtain  all the required  data. This mask
defines the  effective area of  the survey on  the sky, which  is 6437
square  degrees for  the  sample we  use  here.  This  survey area  is
divided into a large number of  smaller subareas for each of which the
NYU-VAGC lists a spectroscopic  completeness $f_{sp}$. This is defined
as the fraction of the  photometrically defined target galaxies in the
subarea for which  usable spectra were obtained. The  average over our
sample galaxies is $\langle f_{sp}\rangle = 0.92$. Within each subarea
the galaxies with spectra can be  assumed to be a random sample of all
possible targets,  with the important exception that  fibres cannot be
closer than  55 arsec in a  single spectroscopic exposure,  so that at
most one  fibre can  be placed  on a galaxy  in a  pair or  group with
smaller angular size  than this. (More fibres may  be assigned to such
clumps if they happen to lie in the overlap region between two or more
spectroscopic  observations.)  It  is  important to  correct for  such
``fibre collisions''  when measuring clustering. As  discussed in more
detail   below,  we   use   the  procedures   of  \citet{Li-06c}   and
\citet{Li-07b} for  this purpose.  These  are based on  comparing pair
counts as a function of angular separation in the spectroscopic sample
and in its parent  photometric sample. General incompleteness is dealt
with  by  weighting  each  galaxy  by $1/f_{sp}$  in  all  statistical
analyses.

A  final observational issue  is that  the SDSS  photometric catalogue
from which our spectroscopic galaxy  sample is drawn is incomplete for
low surface  brightness galaxies \citep{Blanton-05a}.  We discuss this
in    Appendix    B,    based    on    the    recent    analysis    by
\citet{Baldry-Glazebrook-Driver-08}, concluding  that for our purposes
the effects are negligible except  possibly at the very lowest stellar
masses we study.

\subsection{Millennium Simulation, semi-analytic galaxy catalogue, and 
mock redshift surveys}\label{sec:mock}

We have constructed  a set of 20 mock SDSS  galaxy catalogues from the
Millennium Simulation \citep{Springel-05} using  both the sky mask and
the  magnitude and  redshift  limits  of our  real  SDSS sample.   The
Millennium  Simulation uses  $10^{10}$  particles to  follow the  dark
matter distribution in a cubic  region 500$h^{-1}$ Mpc on a side.  The
cosmological      parameters     assumed      are     $\Omega_m=0.25$,
$\Omega_\Lambda=0.75$,  $n=1$,  $\sigma_8=0.9$  and $h=0.73$.   Galaxy
formation within the evolving dark matter distribution is simulated in
postprocessing  using  semi-analytic  methods  tuned to  give  a  good
representation  of the  observed low-redshift  galaxy  population. Our
mock  catalogues   are  based  on   the  galaxy  formation   model  of
\citet{Croton-06}  and  are constructed  from  the publicly  available
$z=0$ data using the methodology of \citet{Li-06c} and \citet{Li-07b}.
These mock catalogues allow us to derive realistic error estimates for
the statistics we measure, including both sampling and cosmic variance
uncertainties.

We   also   use   galaxy   data   from   the   Millennium   Simulation
archive\footnote{http://www.mpa-garching.mpg.de/millennium} to compare
and  contrast  predictions  for   the  total  mass  and  stellar  mass
correlation functions  at $z=0.07$ and at higher  redshift. These data
are based on the galaxy formation model of \citet{DeLucia-Blaizot-07}.

\section{The stellar mass function of galaxies}

\begin{figure}
\centerline{\epsfig{figure=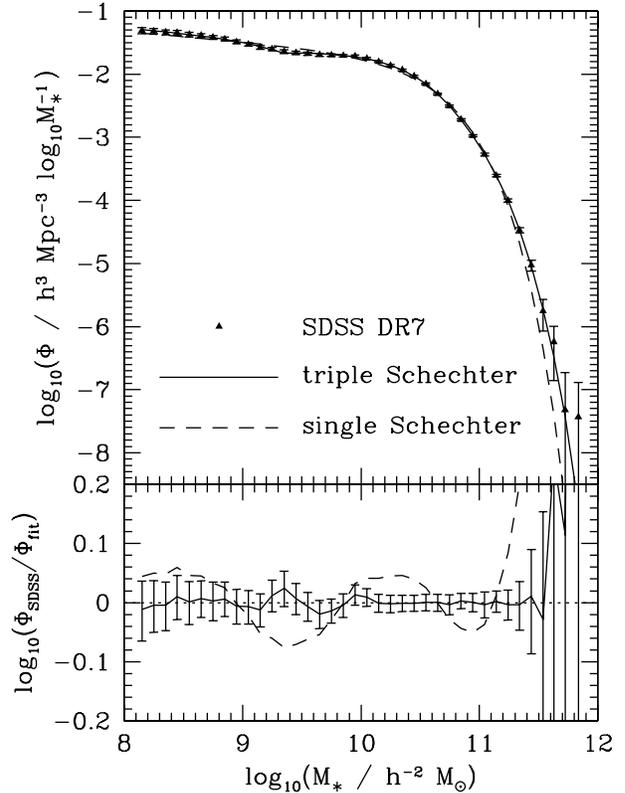,clip=true,width=0.48\textwidth}}
\caption{Stellar mass function for galaxies in the SDSS/DR7 (symbols).
  Error  bars show the  $1\sigma$ scatter  between 20  mock catalogues
  constructed from  the Millennium Simulation using the  same sky mask
  and magnitude and redshift limits as for the real sample. The dashed
  line is the best fit single Schechter function, while the solid line
  is a fit based on three disjoint Schechter functions; its parameters
  are listed  in Table~\ref{tbl:schechter}. The lower  panel shows the
  logarithmic deviation between the data and each of these models. For
  clarity the error bars are  plotted on the more accurate, piece-wise
  fit only.}
\label{fig:smf}
\end{figure}

A  first  statistic  which we  can  estimate  from  the data  we  have
available is the abundance of  galaxies as a function of their stellar
mass. For each observed galaxy  $i$ we define the quantity $z_{max,i}$
to be the maximum redshift  at which the observed galaxy would satisfy
the apparent magnitude limit  of our sample $r\leq 17.6$. Evolutionary
and  K-corrections  are   included  when  calculating  $z_{max,i}$  as
described  by \citet{Li-06a}  and  \citet{Li-07b}. This  allows us  to
define $V_{max,i}$  for the galaxy  in question as the  total comoving
volume of  the survey  out to redshift  $z_{max,i}$. The  stellar mass
function can then be estimated as
\begin{equation}
\Phi(m_*)~\Delta m_* = \sum_i (f_{sp,i}V_{max,i})^{-1},
\end{equation}
where the  sum extends over all  sample galaxies with  stellar mass in
the range $m_* \pm 0.5\Delta m_*$.

In Figure~\ref{fig:smf}, we show  the stellar mass function determined
in  this way  for the  galaxies  in our  sample.  The  error bars  are
estimated from  the scatter  among the mass  functions of our  20 mock
catalogues. As already noted by \citet{Croton-06} the mass function of
this  model agrees reasonably  well with  observation, and  indeed the
number of galaxies in the  real catalogue lies well within the scatter
of the numbers found for the 20 mock catalogues. The error bars should
thus  account correctly  for  the uncertainties  due  to sampling  and
cosmic variance,  but they do not include  systematic uncertainties in
the NYU-VAGC estimates of stellar  masses. Appendix A shows that these
may be reasonably represented by a $\pm 0.1$dex systematic uncertainty
in the  overall mass  scale. Note that  the error bars  on neighboring
points are strongly correlated.

The mass  function of Figure~\ref{fig:smf}  is in good  agreement with
estimates from the 2dFGRS \citep{Cole-01} and from earlier releases of
the      SDSS     \citep[e.g.][]{Bell-03,      Wang-06,     Panter-07,
  Baldry-Glazebrook-Driver-08}.  In Appendix  B we present an explicit
comparison with \citet{Baldry-Glazebrook-Driver-08} which allows us to
assess  the  effect  of   a  small  but  significant  systematic,  the
incompleteness of  the SDSS sample at  low mass {\it  and} low surface
brightness.  This causes  a slight  underestimate of  abundances below
$10^{8.5}M_\odot$  The  purely  statistical  error bars  on  our  mass
function are smaller than in earlier work because of the substantially
larger size  of the  sample. It  is well determined  all the  way from
$10^8M_\odot$  up to  almost $10^{12}M_\odot$.   A fit  with  a single
Schechter function  \citep{Schechter-76} is not  fully consistent with
the data.  In particular, it significantly underpredicts the abundance
of the most massive  galaxies.  Nevertheless, it provides a reasonable
and simple  representation of our results.  The  parameters we obtain,
$\log_{10}(m_*/ h^{-2}M_\odot) = 10.525\pm 0.005$, $\alpha = -1.155\pm
0.008$  and  $\Phi_* =  0.0083\pm  0.0002$  $h^3  {\rm Mpc}^{-3}$  are
similar  to those  found in  earlier  studies.  (The  errors here  are
approximate  $1\sigma$ uncertainties  in  each parameter  marginalised
over the  uncertainties in the  other parameters.)  Our  mass function
data can be considerably better represented by fitting three different
Schechter  functions  (i.e.   with  different parameters)  over  three
disjoint mass  ranges. This representation  is also plotted on  top of
the  data in  Figure~\ref{fig:smf} and  its parameters  are  listed in
Table~\ref{tbl:schechter}. It provides  a compact and accurate summary
of our results.

Another useful representation of the stellar mass function is in terms
of the percentage points  of the cumulative stellar mass distribution.
According to our results, half of  all the stellar mass in galaxies is
in  objects  with individual  stellar  mass  greater than  $1.86\times
10^{10} h^{-2}M_\odot$.  The corresponding 5\%, 10\%, 20\%, 80\%, 90\%
and  95\%  points  are  $1.03$,  $2.47$, $5.86$,  $44.2$,  $65.8$  and
$89.9\times  10^9  h^{-2}M_\odot$  respectively.   (We have  used  our
triple  Schechter fit  to extrapolate  the  low-mass end  of the  mass
function when calculating  these numbers.) Thus 60\% of  all stars are
in galaxies  with stellar mass within  a factor of  2.75 of $1.6\times
10^{10}h^{-2}M_\odot$  which is  about half  the stellar  mass  of the
Milky Way.   A related characteristic  mass which will be  of interest
below  is the  mass-weighted  mean  stellar mass,  which  can also  be
thought of  as the  expected stellar  mass of the  host of  a randomly
chosen star. This is $\bar{M}_\ast = 2.85\times 10^{10} h^{-2}M_\odot$
and is close to the stellar mass of the Milky Way.

A straightforward  integration of our stellar mass  function gives the
mean comoving  stellar mass density  of the low-redshift  Universe (at
$z\sim 0.07$,  see below). This is  $\rho_\ast = 3.14  \pm 0.10 \times
10^8 hM_\odot/{\rm Mpc}^3$, where the  error bar is again derived from
the scatter among our 20  mock catalogues and so accounts for sampling
and cosmic variance effects, but not for systematic errors in the mass
determinations  of individual galaxies.   In the  standard concordance
cosmology \citep[e.g.][]{Komatsu-09} only 3.5\%  of the baryons in the
low-redshift  Universe  are  locked  up  in  stars.   Clearly,  galaxy
formation has been a very inefficient process.

\begin{table*}
\label{tbl:schechter}
\caption{Parameters of a triple  Schechter function fit to the stellar
  mass function of SDSS galaxies}
\begin{center}
\begin{tabular}{cccr} \hline\hline
\multicolumn{1}{c}{mass  range}  &  \multicolumn{1}{c}{$\Phi^\ast$}  &
\multicolumn{1}{c}{$\alpha$}  &  \multicolumn{1}{c}{$\log_{10}M^\ast$}
\\               \multicolumn{1}{c}{($h^{-2}M_\odot$)}               &
\multicolumn{1}{c}{($h^{3}$Mpc$^{-3}\log_{10}M^{-1}$)}                &
\multicolumn{1}{c}{}      &      \multicolumn{1}{c}{($h^{-2}M_\odot$)}
\\  \\  \hline  $8.00<\log_{10}M<9.33$  &  0.0146(5)  &  -1.13(09)  &
9.61(24)  \\  $9.33<\log_{10}M<10.67$   &  0.0132(7)  &  -0.90(04)  &
10.37(02)  \\  $10.67<\log_{10}M<12.00$ &  0.0044(6) &  -1.99(18)  &
10.71(04) \\ \hline
\end{tabular}
\end{center}
\end{table*}

\section{stellar mass correlation functions}

\begin{figure}
\centerline{\epsfig{figure=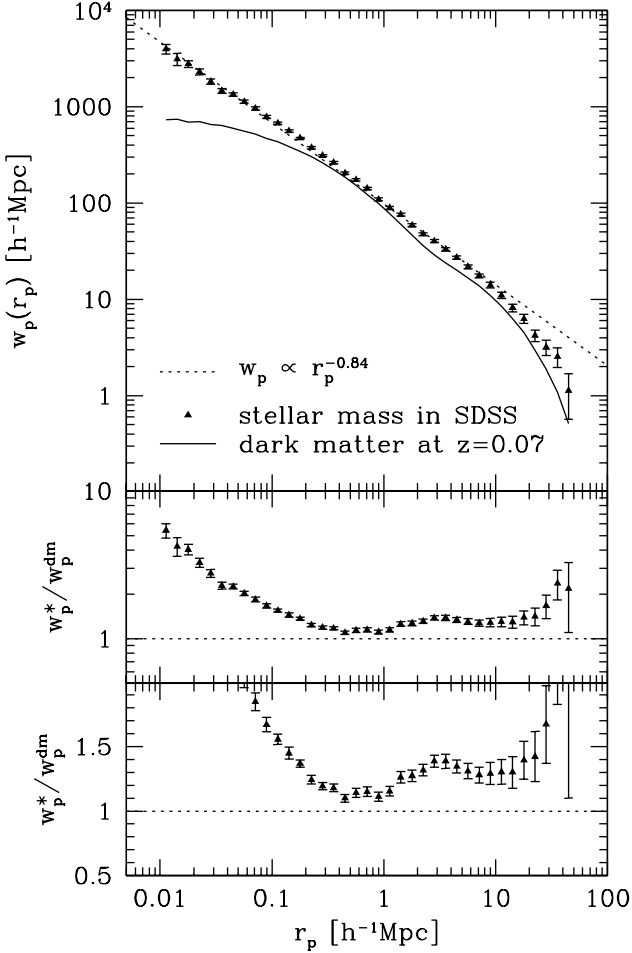,clip=true,width=0.48\textwidth}}
\caption{The  projected stellar mass  autocorrelation function  in the
  SDSS is  plotted as triangles in  the top panel, and  is compared to
  the projected autocorrelation function of dark matter at $z=0.07$ in
  the Millennium  Simulation (the  solid line). The  dashed line  is a
  power-law fit to SDSS data  over the range $10h^{-1}{\rm kpc}< r_p <
  10h^{-1}{\rm   Mpc}$   and   corresponds  to   a   three-dimensional
  autocorrelation function $\xi^\ast(r) = (r/r_0)^{-1.84}$ with $r_0 =
  6.1 h^{-1}{\rm Mpc}$.   The ratio between the stellar  mass and dark
  matter projected autocorrelation  functions is shown logarithmically
  in  the  middle panel  and  linearly  in  the bottom  panel.   Error
  estimates in all three panels  come from the scatter among similarly
  estimated  correlations for  20 mock  galaxy  catalogues constructed
  from the Millennium Simulation  using the same selection criteria as
  the real sample.  See the text for details.}
\label{fig:wrp}
\end{figure}

\begin{figure}
\centerline{\epsfig{figure=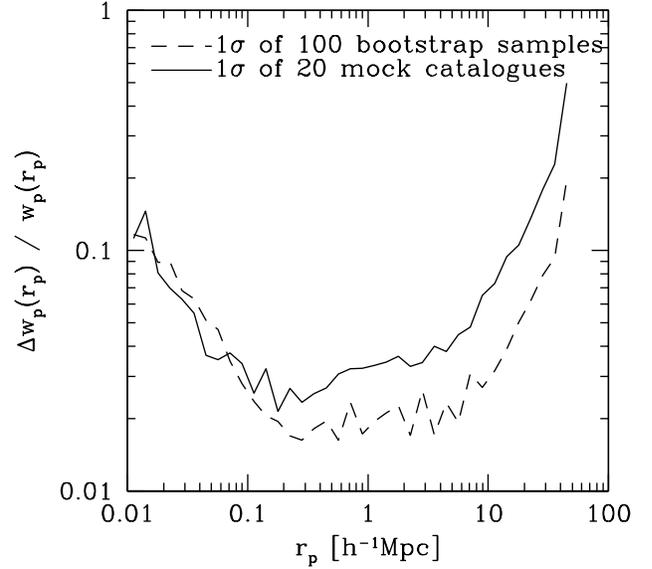,clip=true,width=0.48\textwidth}}
\caption{Relative errors in our estimate of the projected stellar mass
  autocorrelation function based on the scatter between estimates from
  100 bootstrap resamplings of the SDSS data (dashed line) and between
  estimates  from  20  mock  galaxy catalogues  constructed  from  the
  Millennium Simulation  using the same selection criteria  as for the
  real SDSS sample (solid line).}
\label{fig:errors}
\end{figure}

To obtain a  reliable estimate of the clustering  of stellar mass, the
observed sample  must be  compared with a  ``random sample''  which is
unclustered but  fills the same  region of the  sky and has  the same,
stellar mass-dependent redshift  distribution. We construct our random
sample  from the  observed sample  itself, as  described in  detail in
\citet{Li-06a}.  For each real galaxy  we generate 10 sky positions at
random  within  our DR7  mask,  and  we assign  to  each  of them  the
properties of the real galaxy,  in particular, its values of redshift,
stellar mass, $V_{max}$ and  $f_{spec}$. The validity of the resulting
random sample rests on two  requirements: 1) the survey area should be
large  enough that  structures in  the real  sample are  wiped  out by
randomising in  angle; 2) the effective  depth of the  survey must not
vary from  region to region.  Both  are true to good  accuracy for our
sample, which covers $\ga 6000$  deg$^2$, is complete down to $r=17.6$
and  is little  affected by  foreground  dust over  the entire  survey
region. Extensive  tests show that random samples  constructed in this
way produce indistinguishable results from those using the traditional
method \citep{Li-06a}. The advantage  of this technique in the current
application is that it is  guaranteed to maintain the complex relation
between  stellar  mass and  photometric  properties  (and thus  sample
selection criteria) which holds in the real data.

We  begin by estimating  the redshift-space, stellar
mass  correlation function,  $\xi^\ast(r_p,\pi)$ using  an appropriate
version of the \citet{Landy-Szalay-93} estimator:
\begin{equation}
\xi^\ast(r_p,\pi)                                                     =
\frac{DD^\ast(r_p,\pi)-2DR^\ast(r_p,\pi)+RR^\ast(r_p,\pi)}{RR^\ast(r_p,\pi)},
\end{equation}
where  the data-data,  data-random and  random-random pair  counts are
weighted as follows:
\begin{equation}
DD^\ast(r_p,\pi)                    =                   \sum_{(i,j)\in
  DD(r_p,\pi)}\frac{m_{\ast,i}m_{\ast,j}}{f_{coll,ij}f_{spec,i}f_{spec,j}V_{ij}},
\label{eq:DD}
\end{equation}
\begin{equation}
DR^\ast(r_p,\pi)                    =                   \sum_{(i,j)\in
  DR(r_p,\pi)}\frac{m_{\ast,i}m_{\ast,j}}{f_{spec,i}f_{spec,j}V_{ij}},
\end{equation}
\begin{equation}
RR^\ast(r_p,\pi)                    =                   \sum_{(i,j)\in
  RR(r_p,\pi)}\frac{m_{\ast,i}m_{\ast,j}}{f_{spec,i}f_{spec,j}V_{ij}},
\end{equation}
where the sums are over all  pairs $(i,j)$ of the relevant type ($DD$,
$DR$ or  $RR$) with separations in the  two-dimensional separation bin
labelled by  $r_p$ and $\pi$,  the pair separations  perpendicular and
parallel to  the line of  sight. Note that  in the $DD$ and  $RR$ sums
each  pair   appears  twice  (as  $(i,j)$  and   $(j,i)$).   In  these
expressions, $m_{\ast,i}$  and $m_{\ast,j}$ are the  stellar masses of
the  members of  each pair,  $f_{spec,i}$ and  $f_{spec,j}$  are their
associated  spectroscopic completeness fractions,  and $V_{ij}  = {\rm
  min}(V_{max,i},V_{max,j})$  is  the  volume  over which  {\it  both}
galaxies would be  included in the sample.  To  reduce sampling noise,
random samples  are usually constructed with many  more particles than
the real  sample.  To normalise  appropriately, $RR^\ast$ needs  to be
multiplied by $(N_g/N_r)^2$ and $DR^\ast$ by $N_g/N_r$ where $N_g$ and
$N_r$  are the numbers  of galaxies  in the  real and  random samples,
respectively. In our case, $N_r=10\times N_g$.

The final  weight in the  above equations is the  factor $f_{coll,ij}$
which appears in the data-data counts  only. This is a function of the
{\it  angular} separation $\theta_{ij}$  of the  two galaxies,  and is
defined as the fraction of  pairs of angular separation $\theta$ which
are missing from  our spectroscopic sample as a  direct consequence of
the fibre collision problem. We estimate this fraction in the same way
as \citet{Li-06c} and \citet{Li-07b}. We calculate angular correlation
functions $w(\theta)$ for the  spectroscopic sample and for its parent
photometric sample, and we then define
\begin{equation}
f_{coll}(\theta) = [1 + w_{sp}(\theta)]/[1 + w_{ph}(\theta)].
\end{equation}
Detailed tests of this procedure can be found in \citet{Li-06c}.

\begin{figure}
\centerline{\epsfig{figure=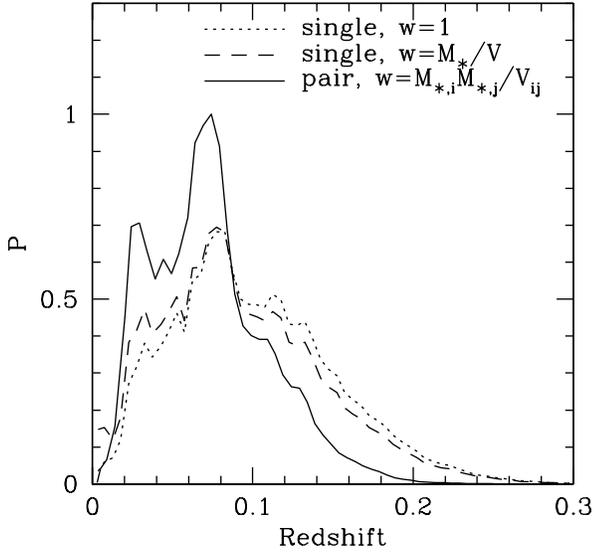,clip=true,width=0.48\textwidth}}
\caption{Redshift distributions  of contributions to  the galaxy count
  (dotted  line), the  cosmic  stellar mass  density estimate  (dashed
  line), and the stellar mass autocorrelation function (solid line) of
  our SDSS  sample.  All curves  are normalised to  have the same
    integral.}
\label{fig:redshift}
\end{figure}

Rather than analysing  these two-dimensional correlation estimates, we
integrate  over  the  line-of-sight  separation  $\pi$  to  obtain  an
estimate   of  the  projected   stellar  mass   correlation  function,
$w_p^\ast(r_p)$.   This  function  is  independent  of  redshift-space
distortions  and is  related by  a  simple integral  transform to  the
three-dimensional  spatial autocorrelation  function for  stellar mass
$\xi^\ast(r)$.  Thus we take
\begin{equation}
w_p^\ast(r_p)=\int_{-\pi_{max}}^{+\pi_{max}}\xi^\ast(r_p,\pi)d\pi=
\sum_i\xi^\ast(r_p,\pi_i)\Delta\pi_i,
\end{equation}
where we choose  $\pi_{max}=40 h^{-1}$ Mpc as the  outer limit for the
depth integration  (in order to limit noise  from distant uncorrelated
regions) so that the summation for computing $w_p^\ast(r_p)$ runs from
$\pi_1 = -39.5 $ h$^{-1}$ Mpc to $\pi_{80} = 39.5$ h$^{-1}$ Mpc, given
that we use bins of width $\Delta\pi_i = 1$ h$^{-1}$ Mpc.

The measurements  of $w_p^\ast(r_p)$ obtained in this  way are plotted
as triangles in the top  panel of Figure~\ref{fig:wrp}. Error bars are
estimated from  the scatter between estimates  of $w_p^\ast(r_p)$ made
by  applying  exactly  the  same   procedures  to  our  20  mock  SDSS
samples. We have also estimated  error bars by bootstrap resampling of
the SDSS  data themselves. As shown  in Figure~\ref{fig:errors}, these
two   methods    produce   consistent   results    on   small   scales
($<200h^{-1}{\rm  kpc}$)  but  at  larger  separations  the  bootstrap
estimates are  consistently smaller than those obtained  from the mock
catalogues. This  is because  the former do  not account  properly for
cosmic variance, which  is the primary source of  uncertainty on large
scales. Over  the range $10h^{-1}{\rm  kpc}< r_p <  10h^{-1}{\rm Mpc}$
the  measurements  have errors  below  10\%  and  are remarkably  well
approximated by  a power law. The  {\it rms} scatter  around the power
law shown  in the  figure is  only 6.9\% over  this range.   At larger
separations $w_p^\ast(r_p)$  rolls off below the  extrapolation of the
power  law.   A  similar  result  with a  slightly  but  significantly
shallower  power  law index  was  obtained  by \citet{Hawkins-03}  for
galaxies (rather than stellar mass) in the 2dF Galaxy Redshift Survey,
and only quite  subtle deviations from power laws  are seen in earlier
SDSS  measurements  of  galaxy correlations  \citep[e.g.][]{Zehavi-04,
  Zehavi-05,  Masjedi-06} This in  turn echoes  the very  first galaxy
correlation              results              obtained              by
\citet{Totsuji-Kihara-69},~\citet{Peebles-74}
and~\citet{Davis-Peebles-83}.

\begin{figure}
\centerline{\epsfig{figure=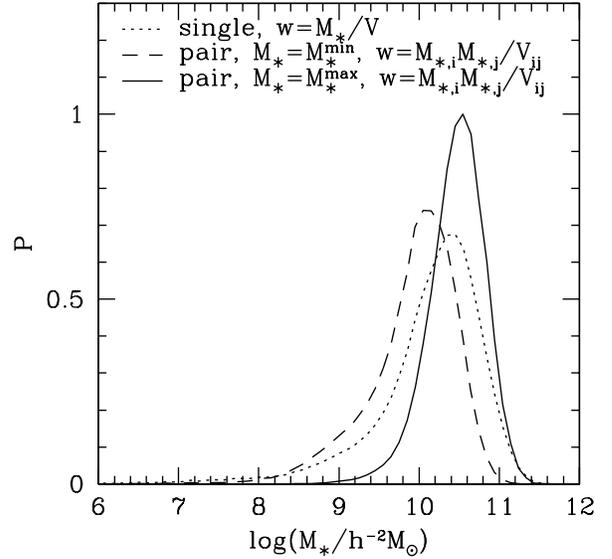,clip=true,width=0.48\textwidth}}
\caption{Distribution across galactic stellar mass of contributions to
  our   estimates  of  cosmic   stellar  mass   density  and   of  the
  autocorrelation function for stellar mass. The dotted line shows the
  contribution to the cosmic stellar mass density coming from galaxies
  in  each  bin   of  $\log  m_\ast$.   The  other   two  curves  show
  contributions to  our estimate  of the stellar  mass autocorrelation
  function  binned as  a  function of  the  stellar mass  of the  most
  massive (solid line)  and of the least massive  (dashed line) galaxy
  in  each pair.  The  three curves  are normalised  to have  the same
  integral.  }
\label{fig:stellarmass}
\end{figure}

For  comparison, Figure~\ref{fig:wrp} also  shows predictions  for the
corresponding  projected   2PCF  for  dark   matter,  $w_p^{dm}(r_p)$,
obtained from the $z=0.065$  snapshot of the Millennium Simulation (as
we  show  below, this  is  the appropriate  mean  depth  for our  SDSS
measurement). A  maximum line-of-sight  depth of 40  $h^{-1}{\rm Mpc}$
was adopted when  computing this statistic in order  to mimic our SDSS
procedures.  The result,  plotted as a solid line in  the top panel of
the  figure, shows  much more  pronounced features  than  stellar mass
correlation function. The ratio between them is shown in the lower two
panels, and  can be thought of  as an estimate  of the scale-dependent
bias between stars  and dark matter.  Our results  are consistent with
bias being independent of scale at $r_p > 1.5h^{-1}{\rm Mpc}$, but the
scale-dependence at smaller separations  is strong, with a total range
of a factor of 5.

In  order to interpret  the stellar  mass autocorrelation  function of
Figure~\ref{fig:wrp} it is helpful to see how the contributions to our
estimate are distributed in  redshift and across galaxies of differing
individual  stellar mass.   Figure~\ref{fig:redshift}  illustrates the
distribution in redshift. The  solid curve histograms contributions to
Equ.~\ref{eq:DD}  from pairs  with  $r_p <  1.0  h^{-1}{\rm Mpc}$  and
$|\pi|<40h^{-1}{\rm  Mpc}$.   The   median  of  this  distribution  is
$z=0.067$  and  the 10  and  90\%  points  are $z=0.025$  and  $0.12$,
respectively.  (These  results are very insensitive  to the particular
$r_p$ range  chosen.) For  comparison, the dotted  line is  a redshift
histogram for all galaxies in  our spectroscopic sample and has median
at $z=0.088$, and 10 and 90\%  points at $z=0.033$ and 0.16, while the
dashed line is  the histogram of contributions to  our estimate of the
cosmic  stellar mass density  (i.e. $m_{\ast,i}/(f_{spec,i}V_{max,i})$
for galaxy $i$ at redshift $z_i$) with median at $z=0.080$, and 10 and
90\% points  at $z=0.026$ and  0.16.  The autocorrelation  function is
dominated by contributions from  a narrower redshift range and peaking
at lower  redshift than  is the parent  sample or the  stellar density
estimate we derived  from it. It is also  remarkable that although our
sample includes almost  half a million galaxies and  covers a sixth of
the sky, the histograms of Figure~\ref{fig:redshift} still show strong
features  which reflect  large-scale structure.   Our  mock catalogues
show that features at this  level, although striking, are expected and
do not appreciably affect our estimate of $w_p^\ast(r_p)$.
 
Figure~\ref{fig:stellarmass}  shows contributions  to our  estimate of
$w_p^\ast(r_p)$ histogrammed as a function  of the stellar mass of the
most  massive (solid  line) and  of  the least  massive (dashed  line)
galaxy in each  pair. As was the case  for the redshift distributions,
we find these curves to be almost independent of the $r_p$ range used;
here   we   again   use   pairs   with  $r_p   <   1.0   h^{-1}$   and
$|\pi|<40h^{-1}{\rm Mpc}$.   For comparison, we also  show a histogram
of contributions to our estimate of the cosmic stellar mass density as
a function  of the stellar mass  of the individual  galaxies.  This is
effectively  the   stellar  mass  function   of  Figure~\ref{fig:smf},
multiplied by  $m_\ast$ and plotted on  a linear scale.   In all three
cases, the dominant contributions  come from a relatively narrow range
of stellar mass centred quite close to the mass of the Milky Way.  The
autocorrelations are  dominated by  contributions from galaxies  in an
even narrower mass range than those dominating the cosmic stellar mass
density, by pairs quite similar to the Local Group.

\section{Discussion}

Our estimates of the mean stellar mass density and of the stellar mass
autocorrelation  function allow  us to  calculate the  average stellar
mass within distance $r$ of a randomly chosen star. This is
\begin{eqnarray}
\label{eq:excessmass}
M_\ast(r) &=& \bar{M}_\ast + 4\pi \rho_\ast \int_0^r
[1+\xi^\ast(r^\prime)]{r^{\prime}}^2dr^\prime \\ \nonumber
 &=& \bar{M}_\ast[1 + (r/355h^{-1}{\rm kpc})^{1.16} + (r/2.8h^{-1}{\rm Mpc})^3],
\end{eqnarray}
where the constant $\bar{M}_\ast=2.85\times 10^{10}h^{-2}M_\odot$, the
mean stellar mass of the chosen star's own galaxy, was calculated from
the stellar  mass function  in section 3.  The second and  third terms
account  for stars  in other  galaxies. It  is interesting  that these
terms  become  comparable  to  $\bar{M}_\ast$ only  at  $350h^{-1}{\rm
  kpc}$.   This is  at  least thirty  times  the size  of the  stellar
component  of a  typical galaxy,  a factor  which quantifies  how much
dissipative effects have condensed  the visible components of galaxies
with respect to the larger scale dissipationless hierarchy.

It is striking that our measured stellar-mass autocorrelation function
is very well  fit by a power law over about  three orders of magnitude
in spatial  scale. This behaviour breaks down  dramatically on smaller
scales where ``one-galaxy'' contributions  cause $\xi^\ast$ to jump in
amplitude   by   about  two   orders   of   magnitude  (see   equation
(\ref{eq:excessmass})) but  also on larger scales where  we detect the
roll-down in  amplitude expected  in $\Lambda$CDM universes.   When it
was first  seen in  galaxy autocorrelations, such  power-law behaviour
was interpreted as evidence  for the scale-free nature of hierarchical
clustering  under  gravity  \citep[e.g.][]{Peebles-80}.   High-quality
numerical simulations have improved  our understanding of this process
considerably, showing  that precise power-law behaviour  should not be
expected,  either on  highly  nonlinear scales  or  in the  transition
between  linear and nonlinear  scales.  Fig.~\ref{fig:wrp}  shows dark
matter   correlations   in  the   Millennium   Simulation  to   depart
substantially from  power-law behaviour in  both these regimes.  It is
thus surprising  that the observed stellar mass  autocorrelation is an
excellent power  law between  10$h^{-1}$kpc and 10$h^{-1}$Mpc.  In our
standard  structure   formation  model,  this   must  be  seen   as  a
coincidence.  Different processes  are required  to  cause convergence
towards the observed power law on different scales.
 
\begin{figure}
\centerline{\epsfig{figure=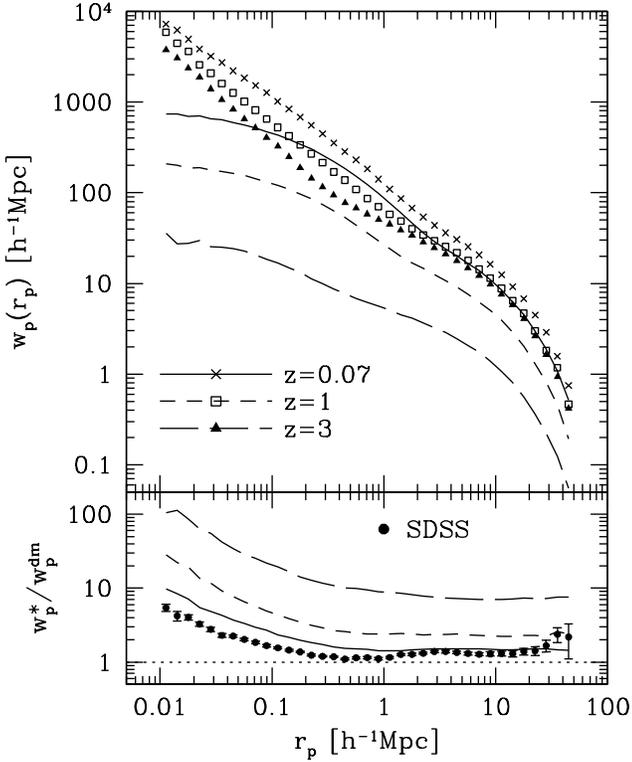,clip=true,width=0.48\textwidth}}
\caption{Projected autocorrelation functions at $z= 0.07, 1.0$ and 3.0
  for  the stellar  and for  the dark  matter mass  in  the Millennium
  Simulation.   The semi-analytic model  of \citet{DeLucia-Blaizot-07}
  is used to  specify the positions, velocities and  stellar masses of
  the galaxies.  The upper  panel shows results  for the  stellar mass
  (symbols) and for  the dark matter (lines) separately.  Lines in the
  lower panel  show the ``bias'', i.e.  the ratio at  each redshift of
  the stellar mass  and dark matter functions in  the upper panel. The
  symbols  in  the  lower  panel   repeat  the  SDSS  bias  data  from
  Figure~\ref{fig:wrp}  and should  be compared  with the  model curve
  shown as a solid line.}
\label{fig:high_z}
\end{figure}

We  use the  Millennium Simulation  to explore  this issue  further in
Figure~\ref{fig:high_z}.  This  compares  projected stellar  and  dark
matter autocorrelation  functions at $z=0.07,  1.0$ and 3.0  using the
semi-analytic  model  of  \citet{DeLucia-Blaizot-07}  to  specify  the
stellar masses  of the ``galaxies''.   The upper panel shows  the dark
and stellar  mass autocorrelations  separately, while the  lower panel
shows their  ratio, the ``bias factor'',  as a function  of scale. The
well-known  result that  galaxy correlations  are predicted  to evolve
much more weakly  than dark matter correlations is  very clear in this
plot. More  interesting in  the current context  is the fact  that the
predicted $z=0$ stellar  mass correlations are much closer  to a power
law for $r_p<10h^{-1}$Mpc than are those for the dark matter, although
the  deviations are  significantly larger  than  in the  real data  of
Fig.~\ref{fig:wrp}.  The two  can be  compared in  the lower  panel of
Fig.~\ref{fig:high_z} where  the SDSS bias  data are shown  as symbols
with error  bars. Although the model reproduces  the ``two-halo'' part
of the  observed function very well, it  overpredicts the ``one-halo''
part by about 50\% and this leads to the bulge above a power law which
is visible in the upper  panel.  At higher redshifts the model stellar
mass  autocorrelations  maintain their  power-law  behaviour on  small
scales, becoming even steeper than  at $z=0$.  By $z=3$ the transition
between  one  and  two  halo  terms  has  become  very  obvious.  This
reinforces  the conclusion that  the remarkably  precise power  law of
Fig.~\ref{fig:wrp} is just a coincidence.

\begin{figure}
\centerline{\epsfig{figure=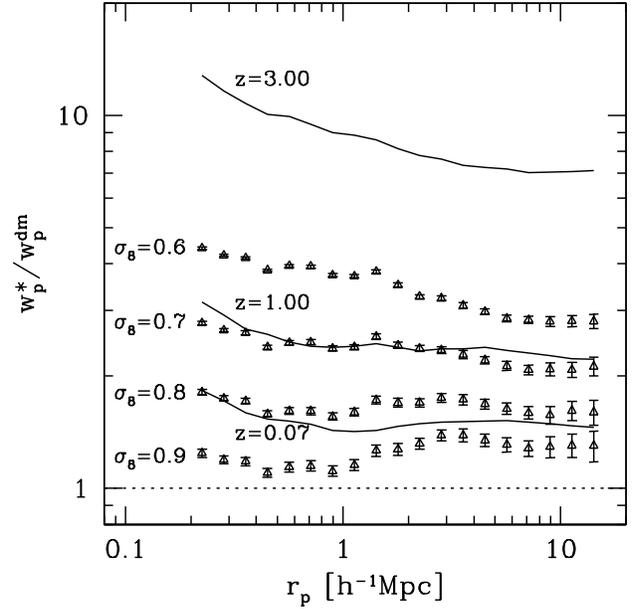,clip=true,width=0.48\textwidth}}
\caption{Bias  as  a function  of  scale  for  the SDSS  stellar  mass
  distribution taking $z=0.07$  dark matter correlations corresponding
  (approximately)   to  the   Millennium  Simulation   cosmology  with
  $\sigma_8=0.9,  0.8,  0.7$  and  0.6  (triangles  with  error  bars,
  labelled by $\sigma_8$). These  are compared to model bias functions
  at  $z=0.07, 1$  and  3 for  the  stellar mass  distribution in  the
  \citet{DeLucia-Blaizot-07}  galaxy   formation  model  (solid  lines
  labelled by redshift). Only  for $\sigma_8=0.8$ is the bias function
  inferred from the SDSS data as flat and as free from features at the
  1-halo$/$2-halo transition as the model functions.}
\label{fig:sigma8}
\end{figure}

Another   interesting   feature   of   the  model   bias   curves   in
Fig.~\ref{fig:high_z}  is the  fact that  they are  smooth  and almost
constant   over   the   range   $200h^{-1}{\rm   kpc}<r_p<15h^{-1}{\rm
  Mpc}$.  They show  at most  a very  weak feature  at  the transition
between  the 1-halo  and 2-halo  regimes,  despite the  fact that  the
projected correlation functions from which they were derived show such
features quite  clearly. In contrast,  when we divide  our featureless
observed  $w_p^\ast(r_p)$ by the  corresponding $z=0.07$  function for
dark  matter in the  Millennium Simulation,  the resulting  bias curve
shows an  obvious step at the 1-halo/2-halo  transition which reflects
the marked  change in  slope of the  dark matter correlations  at this
point (see the bottom panel of Fig.~\ref{fig:wrp}). This suggests that
the  amplitude of  mass fluctuations  is  too high  in the  Millennium
Simulation, and that  the {\it shape} of our  stellar mass correlation
function  can  be  used  to  estimate the  value  of  the  fluctuation
amplitude parameter $\sigma_8$.

We illustrate  this possibility in Fig.~\ref{fig:sigma8}  where we use
the  projected dark  matter  correlations measured  in the  Millennium
Simulation  at redshifts  of 0.32,  0.62 and  0.99 to  represent those
expected at  $z=0.07$ in cosmologies with  identical parameters except
that    $\sigma_8$    is    reduced    to    0.8,    0.7    and    0.6
respectively\footnote{This  scaling assumes  that  the nonlinear  dark
  matter  power spectrum  depends on  the  shape of  the linear  power
  spectrum and its extrapolated amplitude, but not on other parameters
  such as $\Omega_m$ or  $\Omega_\Lambda$. This approximation is quite
  good over the  range of scales relevant here,  and is certainly good
  enough  for  the present,  essentially  qualitative argument.}.   We
derive bias  functions from our  SDSS stellar mass  correlations using
these three functions  in addition to the original  $z=0.07$ data, and
we compare their  shapes to those of model  bias functions at $z=0.07,
1$  and 3  taken from  Fig.~\ref{fig:high_z}.  For  $\sigma_8=0.9$ the
1-halo/2-halo transition feature is clear  and is much larger than any
feature  seen  in  the  models.  For $\sigma_8=0.7$  the  feature  has
reversed sign  and the inferred bias  function has a  much more marked
overall  slope than  any of  the models.  The  ``Goldilocks'' solution
appears to be close to $\sigma_8=0.8$ where the bias function is quite
flat and the transition feature  is almost absent.  This accords quite
well  with the  values of  $\sigma_8$ suggested  by joint  analysis of
WMAP5 and low-redshift supernova and BAO data \citep{Komatsu-09}.

If  it can  be  shown that  the  smooth behaviour  of  the model  bias
functions  of  Fig.~\ref{fig:sigma8}  is  generic  to  all  physically
reasonable  $\Lambda$CDM galaxy formation  models, then  these results
suggest   a  powerful  cosmological   test.  Projected   stellar  mass
correlation functions can be estimated  as a function of redshift from
any large  galaxy survey with  sufficiently good photometry  to obtain
robust   and   relatively    precise   photometric   redshifts.    The
1-halo/2-halo   feature  typically   occurs  at   radii   where  these
correlation  function estimates have  their best  signal-to-noise, and
thus  is much  easier to  measure  than, say,  the baryon  oscillation
feature. Requiring  the bias  functions to be  smooth and  nearly flat
will  determine  $\sigma_8(z)$.  In  addition,  the  location  of  the
transition feature in the  individual correlation functions provides a
length-scale which can be used to get an angular size distance to each
redshift. If it works. this  scheme thus provides measurements of {\it
  both} the  functions of redshift  which are needed to  constrain the
nature of Dark Energy.

Although  the \citet{DeLucia-Blaizot-07}  galaxy catalogue  provides a
surprisingly good fit  to the observed clustering of  stellar mass, it
does  much less well  when compared  to the  stellar mass  function of
Fig.~\ref{fig:smf}.  The very small error bars highlight discrepancies
which could be ignored when comparing to previously published and less
precise  mass functions.  The  most important  of  these concerns  the
abundances at  low mass, where the  SDSS measurements are  a factor of
two below  the galaxy formation  models implemented on  the Millennium
Simulation  over the  full range  $8 <  \log M_*/M_\odot  <  9.5$ (see
Fig.~\ref{fig:baldry}  below). Clearly,  galaxy formation  in low-mass
halos was considerably less efficient  in the real Universe than these
models predict.   Precise statistics  for the distribution  of stellar
mass of the  kind obtained in this paper  provide hard constraints for
galaxy formation  models, and  learning what is  required to  fit them
properly may teach us much about the physics of galaxy formation.

\section*{Acknowledgments}
We thank the referee for helpful comments and Michael Blanton for help
in  understanding  the stellar  mass  estimates  in  NYU-VAGC.  CL  is
supported  by  the   Joint  Postdoctoral  Programme  in  Astrophysical
Cosmology  of  Max  Planck  Institute for  Astrophysics  and  Shanghai
Astronomical Observatory, by NSFC (10533030, 10633020), by 973 Program
(No.2007CB815402)  and  by the  Knowledge  Innovation  Program of  CAS
(No.KJCX2-YW-T05).

Funding  for the  SDSS and  SDSS-II has  been provided  by  the Alfred
P.  Sloan  Foundation, the  Participating  Institutions, the  National
Science  Foundation,  the  U.S.  Department of  Energy,  the  National
Aeronautics and Space Administration, the Japanese Monbukagakusho, the
Max  Planck Society,  and  the Higher  Education  Funding Council  for
England. The SDSS Web Site is http://www.sdss.org/.

The SDSS is  managed by the Astrophysical Research  Consortium for the
Participating  Institutions. The  Participating  Institutions are  the
American Museum  of Natural History,  Astrophysical Institute Potsdam,
University  of Basel,  University of  Cambridge, Case  Western Reserve
University,  University of Chicago,  Drexel University,  Fermilab, the
Institute  for Advanced  Study, the  Japan Participation  Group, Johns
Hopkins University, the Joint  Institute for Nuclear Astrophysics, the
Kavli Institute  for Particle  Astrophysics and Cosmology,  the Korean
Scientist Group, the Chinese  Academy of Sciences (LAMOST), Los Alamos
National  Laboratory, the  Max-Planck-Institute for  Astronomy (MPIA),
the  Max-Planck-Institute  for Astrophysics  (MPA),  New Mexico  State
University,   Ohio  State   University,   University  of   Pittsburgh,
University  of  Portsmouth, Princeton  University,  the United  States
Naval Observatory, and the University of Washington.

\bibliography{ref}

\appendix

\section{Systematic biases due to the stellar mass definition}

\begin{figure}
\vspace{0.2cm}
\centerline{\psfig{figure=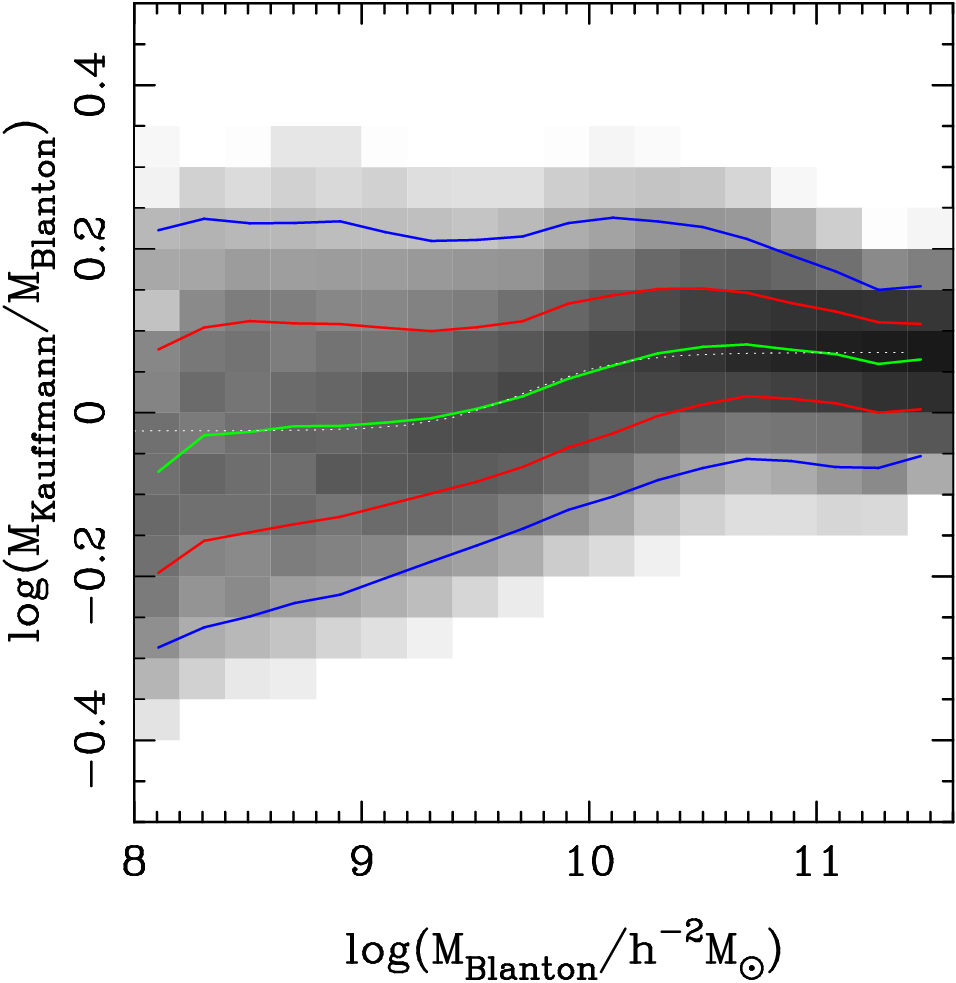,width=0.43\textwidth}}
\vspace{0.5cm}
\centerline{\psfig{figure=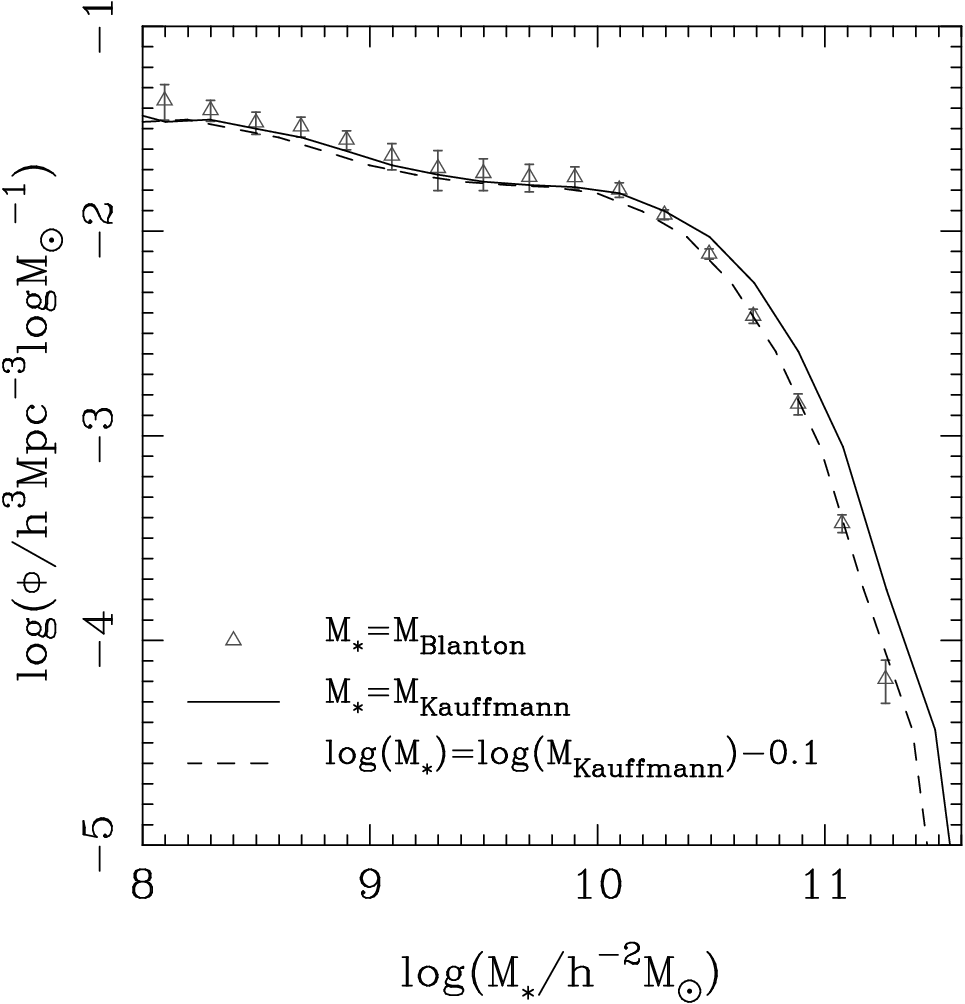,width=0.43\textwidth}}
\caption{{\it     Upper:}      Stellar     mass     estimates     from
  \citet{Blanton-Roweis-07}, $M_{\rm  Blanton}$, compared to  those of
  \citet{Kauffmann-03}, $M_{\rm Kauffmann}$,  as a function of $M_{\rm
    Blanton}$.   The  gray  scale   is  the  distribution  of  $M_{\rm
    Kauffmann}/M_{\rm   Blanton}$    at   each   value    of   $M_{\rm
    Blanton}$.  The solid  lines (from  bottom to  top) are  the 10\%,
  25\%,  50\%, 75\%,  and  90\% quantiles  of  this distribution.  The
  dotted line  is a  fit to  the median (see  the text).  {\it Lower:}
  Stellar  mass functions  for DR4  estimated using  $M_{\rm Blanton}$
  (symbols) and  $M_{\rm Kauffmann}$ (solid line). The  dashed line is
  obtained from the solid line by shifting it by $\Delta\log M = -0.1$
}
\label{fig:gamk.vs.mike}
\end{figure}

\begin{figure}
\vspace{0.2cm}
\centerline{\psfig{figure=fA2a.ps,width=0.43\textwidth}}
\vspace{0.5cm}
\centerline{\psfig{figure=fA2b.ps,width=0.43\textwidth}}
\caption{{\it     Upper:}      Stellar     mass     estimates     from
  \citet{Blanton-Roweis-07}, $M_{\rm Blanton}$, compared to $(r-i)$ --
  $r$ based  estimates from  the formulae of  \citet{Bell-03}, $M_{\rm
    Bell}$, as a function of  $M_{\rm Blanton}$. The gray scale is the
  distribution  of $M_{\rm  Bell}/M_{\rm  Blanton}$ at  each value  of
  $M_{\rm  Blanton}$. The  solid lines  (from bottom  to top)  are the
  10\%, 25\%, 50\%, 75\%, and 90\% quantiles. The dotted line is a fit
  to the median  (see the text).  {\it Lower:}  Stellar mass functions
  for  DR7 estimated  using  $M_{\rm Blanton}$  (symbols) and  $M_{\rm
    Bell}$ (solid  line). The dashed  line is obtained from  the solid
  line by by shifting it by $\Delta\log M = -0.05$ }
\label{fig:bell.vs.mike}
\end{figure}

\begin{figure}
\centerline{\psfig{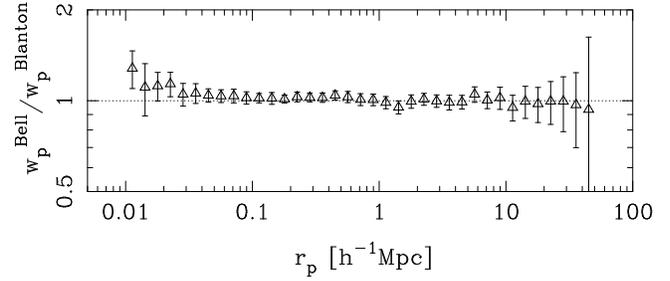}}
\caption{Ratio of the  projected stellar mass autocorrelation function
  in the SDSS  estimated using $M_{\rm Bell}$ to  that estimated using
  $M_{\rm Blanton}$.}
\label{fig:wrp_bell.vs.mike}
\end{figure}

In order  to explore  possible systematics in  our results due  to the
particular  \citet{Blanton-Roweis-07} stellar  mass estimates  we have
used,  we   here  repeat  parts   of  our  analysis  using   both  the
spectroscopy-photometry-based stellar  masses of \citet{Kauffmann-03},
$M_{\rm  Kauffmann}$, and  the simpler  colour-magnitude-based stellar
masses of \citet{Bell-03}, $M_{\rm Bell}$.

For $M_{\rm Kauffmann}$ we have had to go back to the {\tt Sample dr4}
of   NYU-VAGC,    which   is   based   on   SDSS    Data   Release   4
\citep[DR4;][]{Adelman-McCarthy-06}, since  $M_{\rm Kauffmann}$ is not
available for later releases. From {\tt Sample dr4} we have selected a
sample   of  300,596   galaxies  using   the  same   criteria   as  in
\S~\ref{sec:sample}.   Measurements of  $M_{\rm Kauffmann}$  are taken
from            the             MPA/JHU            SDSS            DR4
database\footnote{http://www.mpa-garching.mpg.de/SDSS/DR4/}.        The
reader is referred to  \citet{Kauffmann-03} for a detailed description
of the methodology used to  derive $M_{\rm Kauffmann}$.  In brief, the
amplitude of  the 4000-\AA\ break  $D_{4000}$ and the strength  of the
H$\delta$  absorption  line  were  obtained  from  stellar  population
synthesis  models  for a  library  of  32,000  diverse star  formation
histories.  A  maximum-likelihood  estimate  of the  $z$-band  stellar
mass-to-light ratio  can then  be estimated for  each galaxy  from its
observed $D_{4000}$ and H$\delta$  indices and applied to its observed
$z$-band luminosity to estimate its stellar mass.

The   upper    panel   of   Figure~\ref{fig:gamk.vs.mike}    shows   a
galaxy-by-galaxy   comparison  of   $M_{\rm  Kauffmann}$   to  $M_{\rm
  Blanton}$, the stellar  mass estimate used in the  main part of this
paper,  as a  function  of  $M_{\rm Blanton}$.   As  already shown  in
\citealt{Blanton-Roweis-07} (see their Fig.~17), the two estimates are
very similar, with a typical scatter of 0.1 dex and with offsets below
$0.1$  dex  at all  $M_{\rm  Blanton}$.   The  median of  $\log(M_{\rm
  Kauffmann}/M_{\rm   Blanton})$,  $\Delta  m$,   as  a   function  of
$m\equiv\log(M_{\rm  Blanton}/h^{-2}M_\odot)$ can be  well represented
by a hyperbolic tangent function,
\begin{equation}\label{eqn:gamk.vs.mike}
\Delta m=0.0256+0.0478\tanh[(m-9.73)/0.417],
\end{equation}
which we show as a dotted line in the figure.

In the lower panel  of Figure~\ref{fig:gamk.vs.mike}, the stellar mass
function  estimated from  $M_{\rm Kauffmann}$  is plotted  as  a solid
line,  and  is  compared  to  that from  $M_{\rm  Blanton}$  shown  as
triangles.  Both estimates  are based on DR4 so  the $M_{\rm Blanton}$
function differs from that  of Figure~\ref{fig:smf}. Error bars on the
latter  show   the  $1\sigma$  scatter  between   20  mock  catalogues
constructed from  the Millennium Simulation  using the same  sky mask,
magnitude and redshift limits as for the real DR4 sample, as described
in \S~\ref{sec:mock}.  The two mass functions are consistent with each
other at masses below $\sim 5\times10^{10} M_\odot$. At higher masses,
$M_{\rm  Kauffmann}$-based  mass function  lies  above  that based  on
$M_{\rm Blanton}$.  This difference almost disappears if the former is
shifted to the left  by $\Delta \log M = -0.1$ as  shown by the dashed
line. With this small shift in mass scale, the difference in abundance
between the two determinations nowhere exceeds 0.1 dex.

We now  consider $M_{\rm Bell}$.  Here we use  the DR7 data as  in the
main text and we compute $M_{\rm Bell}$ for each galaxy from its $r-i$
colour  and  its  $r$-band  luminosity  using the  formulae  given  by
\citet[][see their Appendix  2 and Table 7]{Bell-03} for  a Kroupa IMF
\citep{Kroupa-01}.   This is the  IMF adopted  by \citet{Kauffmann-03}
and   it  is   quite  similar   to   the  Chabrier   IMF  assumed   by
\citet{Blanton-Roweis-07}.

Figure~\ref{fig:bell.vs.mike}    is    identical    in    format    to
Figure~\ref{fig:gamk.vs.mike}.   The   upper   panel   is   a   direct
galaxy-by-galaxy comparison  of $M_{\rm Bell}$  and $M_{\rm Blanton}$,
while the  lower panel compares  DR7 mass functions obtained  with the
two stellar mass estimates The symbols and lines have the same meaning
as in the previous figure.  The median mass ratio $M_{\rm Bell}/M_{\rm
  Blanton}$ is almost independent of $M_{\rm Blanton}$ at masses above
$\sim 10^{10}h^{-2}M_\odot$,  but it increases  substantially at lower
masses, reaching  0.3 dex at $10^8h^{-2}M_\odot$.   This behaviour can
be modelled by a quartic function,
\begin{equation}
\Delta m=2.0-0.043m-0.045m^2+0.0032m^3-2.1\times10^{-5}x^4,
\end{equation}
where    $m=\log(M_{\rm     Blanton}/h^{-2}M_\odot)$    and    $\Delta
m=\log(M_{\rm Bell}/M_{\rm Blanton})$.  With  a scale shift of $\Delta
\log M = -0.05$, the stellar  mass function based on $M_{\rm Bell}$ is
a  good match  to  that based  on  $M_{\rm Blanton}$  at masses  above
$10^{10}h^{-2}M_\odot$, but is high by $\sim 0.1$ dex at lower masses.

In summary,  differences between these  three observational estimators
of  stellar  mass affect  our  mass  function determination  primarily
through  small off-sets in  the mass  scale. Once  this is  taken into
account, abundance  offsets between the different  estimates are quite
small     across    the     full    range     $10^8h^{-2}M_\odot<    M
<10^{11.5}h^{-2}M_\odot$  that  we consider.  As  we now  demonstrate,
effects on our stellar mass autocorrelation function estimate are much
smaller.

In Figure~\ref{fig:wrp_bell.vs.mike} we plot  the ratio of the stellar
mass autocorrelation  function computed  using $M_{\rm Bell}$  to that
computed using $M_{\rm Blanton}$.  The two $w_p(r_p)$ measurements are
indistinguishable on scales above about  30 kpc. For $r_p<30$ kpc, the
amplitude of the $M_{\rm Bell}$-based correlation function is slightly
higher, but still within the  error bars of the measurements. This can
be    understood    from    the     fact    that,    as    shown    in
Figure~\ref{fig:stellarmass}, the  mass autocorrelations are dominated
by contributions from galaxies in a relatively narrow mass range where
Figures~\ref{fig:gamk.vs.mike}   and~\ref{fig:bell.vs.mike}  show  the
scatter between the various estimators  to be small. Any scale off-set
between  them  drops out  in  the  definition  of the  autocorrelation
function.

\section{Incompleteness for low surface brightness galaxies}

\begin{figure}
\centerline{\psfig{figure=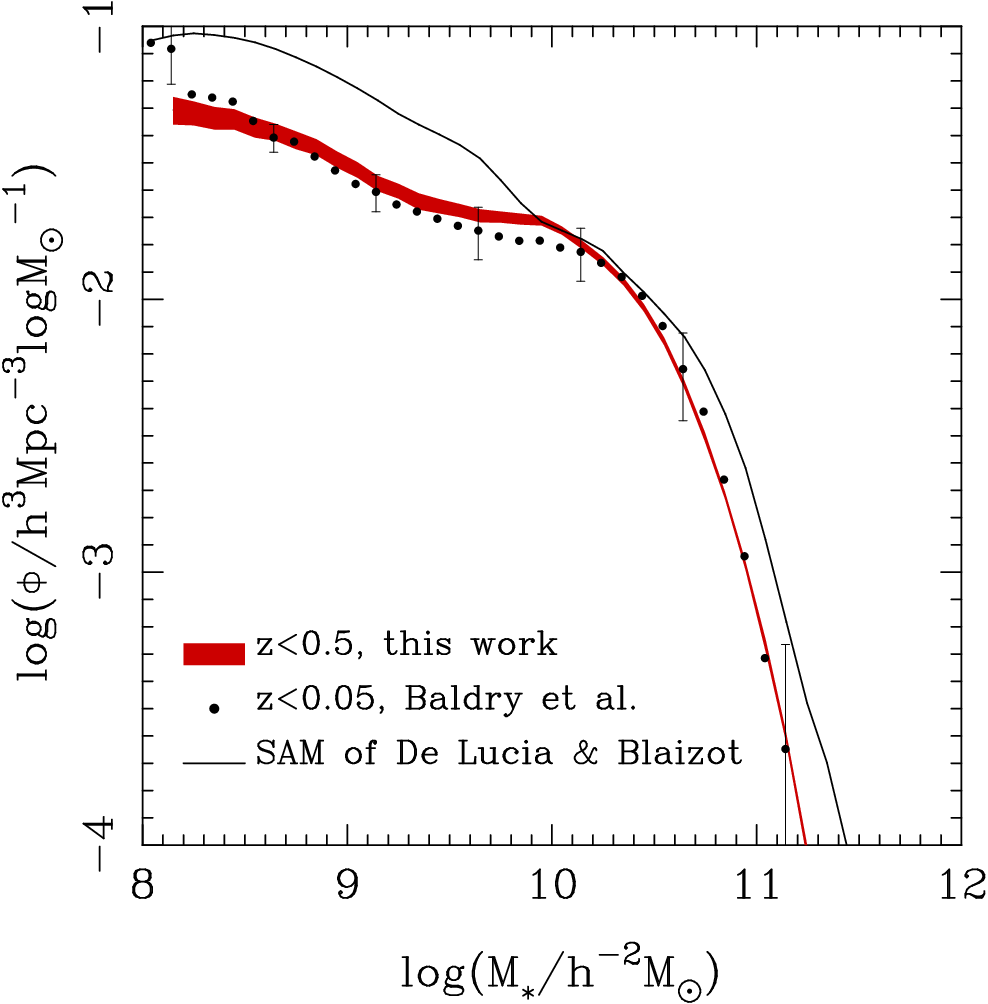,width=0.43\textwidth}}
\caption{The  1-$\sigma$ confidence  region for  our DR7  stellar mass
  function (median redshift 0.09) is plotted as a shaded region and is
  compared to the  stellar mass function for DR4  galaxies at $z<0.05$
  presented     by     \citet{Baldry-Glazebrook-Driver-08}     (filled
  circles).  The  stellar mass  function  for  $z=0$  galaxies in  the
  Millennium     Simulation      galaxy     formation     model     of
  \citet{DeLucia-Blaizot-07} is also shown as a solid line. Error bars
  on some  of the circles show  the $1\sigma$ scatter  between 20 mock
  catalogues constructed from the Millennium Simulation using the same
  sky  mask,  magnitude  and  redshift  limits as  in  the  sample  of
  \citet{Baldry-Glazebrook-Driver-08}.  Clearly   the  two  SDSS  mass
  functions agree well.}
\label{fig:baldry}
\end{figure}

As  discussed  by  \citet{Blanton-05a}  the SDSS  galaxy  samples  are
incomplete  for low surface  brightness (SB)  objects. This  affects a
negligible fraction  of high-mass galaxies but can  become serious for
dwarfs.  Thus  one may  be concerned that  our stellar  mass functions
underestimate     the      abundance     of     low-mass     galaxies.
\citet{Baldry-Glazebrook-Driver-08} have  examined this issue  in some
detail, deriving the bivariate distribution of SDSS galaxies in SB and
stellar mass. They found that for the stellar masses studied here, the
distribution of  $\log({\rm SB})$  is approximately gaussian  at given
stellar mass, with a mean which decreases linearly with $\log(M_\ast)$
over the  range $10^{8.5}$  to $10^{11} M_\odot$  and a  scatter which
increases slightly towards lower  masses (see their Fig.4). As stellar
mass  decreases, the  fraction  of  galaxies with  SB  below the  SDSS
completeness limit thus steadily increases. Nevertheless, at the lower
limit  to which we  plot our  mass function  ($\log(M_\ast) =  8.3$ in
Fig.~4   of  \citet{Baldry-Glazebrook-Driver-08}   since   they  adopt
$h=0.7$)  the estimated  completeness  is still  well  above 70\%.  We
therefore expect incompleteness to have  at most a minor effect on our
results.

In Figure~\ref{fig:baldry}  we compare  our DR7 stellar  mass function
explicitly   to  the  one   which  \citet{Baldry-Glazebrook-Driver-08}
estimated           for            DR4           galaxies           at
$z<0.05$.  \citet{Baldry-Glazebrook-Driver-08}   used  stellar  masses
estimated as the average of those obtained by four different published
methods      \citep{Kauffmann-03,      Glazebrook-04,     Gallazzi-05,
  Panter-07}. In the mean this should  give a mass scale close to that
of  \citet{Blanton-Roweis-07}. Because  of  the substantially  smaller
volume   of  their  sample,   the  \citet{Baldry-Glazebrook-Driver-08}
results are substantially  noisier than our own. (In  order to include
cosmic variance effects, we have  derived the error bars in the figure
from mock catalogues  in a similar way to those shown  on our own mass
functions in the main body of  our paper.) Agreement is very good over
the full mass range plotted.  Furthermore, the analysis in their paper
shows that SB incompleteness affects only the two or three lowest mass
points plotted in Figure~\ref{fig:baldry} and so is negligible for our
purposes.  This  figure also  shows the stellar  mass function  of the
Millennium      Simulation     galaxy      formation      model     of
\citet{DeLucia-Blaizot-07}.   This agrees  well with  the observations
around the knee  of the function, but it predicts  the rarest and most
massive  galaxies to  have stellar  masses 0.2  dex larger  than those
estimated in  SDSS, and, as noted  in the final paragraph  of our main
text,  it  predicts  an   abundance  of  low-mass  galaxies  which  is
substantially larger than observed.

Finally we  note that SB incompleteness  has no effect  on the stellar
mass  autocorrelations  that  we  estimate  because,  as  we  show  in
Figure~\ref{fig:stellarmass},  these  are  dominated by  contributions
from galaxies of substantially larger stellar mass.

\bsp
\label{lastpage}

\end{document}